Title
Low-energy Coincidences in Cosmic Ray Detection.

Author

Jan Oldenziel    Former staff member of the Faculty of Experimental Physics, University of Amsterdam
                 Guest collaborator in Hisparc Group, Nikhef National Institute for Subatomic Physics
                 Science Park 105, 1098 XG, Amsterdam, The Netherlands
                 e-mail: janolden@nikhef.nl, jgoldenziel@xs4all.nl

Abstract

We measured signals originating from cosmic rays, using two rectangular-block scintillation detectors at various positions. The signals were analyzed by a slightly modified signal analyzer from project 'MuonLab',  designed to measure the lifetime and velocity of muons from cosmic radiation in high school education. In our experiment we focused on the possibilities of the apparatus to measure the time difference 'deltatime (DT)' between signals in two scintillation detectors and the signal amplitude 'pulse height (PH)' of the signal in one detector.

We performed measurements in two arrangements: first vertical, detectors parallel and above each other, second horizontal, detectors parallel and next to each other. We observed that in both the vertical and horizontal arrangements, low-energy signals from cosmic rays in the two detectors showed unexpected coincidences.  Under normal experimental conditions usually these effects are not observed because the amplification in the photomultiplier (PMT) as well as the trigger level are intentionally chosen to avoid noise and excessively high count rates. The rate at which the coincidences occur is very  small, more than a thousand times smaller than the count rates of the individual detectors. The measurement of the deltatime spectrum, however, appears to be a sensitive way to detect the presence of these rare and unexpected effects.

The time differences between the signals at the two analyzer inputs were measured with a resolution of 0.5 ns and the pulse height measurement had a resolution of approximately 8 mV. It is estimated that the low-amplitude signals involved in the coincidences must correspond to energy absorptions less than 100 keV in the scintillator material.

1. Introduction

Cosmic radiation has been widely studied since its discovery in the beginning of the 20th century. In the Netherlands a Hisparc project[1] has been set up, which is also suitable for educational use by high school students. The Hisparc project at the Nikhef Institute of Subatomic Physics in Amsterdam was recently described extensively by Kasper van Dam et al.[3].

Next to the Hisparc project, a MuonLab project[2] has been developed within the Nikhef. It offers the possibility of measuring the lifetime and the velocity of muons from cosmic radiation. It is also suitable for Rossi's experiments to show the production of secondary particles in material above two detectors by coincidence measurement.

The experiments within the MuonLab project may be used for an educational project for high school students or bachelor's degrees in physics at universities. The MuonLab apparatus, designed by the Nikhef Institute for Subatomic Physics, has been sold at cost price to schools and academic institutes in various countries. The apparatus is still available at cost price at the Nikhef Institute for Subatomic Physics.

The observation of the energy spectrum of random coincidences in cosmic ray measurements, as measured in the Hisparc project, suggested that a more elaborate study of coincidences at low energies using the MuonLab apparatus might be interesting.

## 2. Method

The MuonLab instrument we have used, described in more detail in the website https://www.nikhef.nl/muonlab, basically consists of two scintillation detectors connected to a signal analyzer interfaced to a Linux computer.

Schematically:

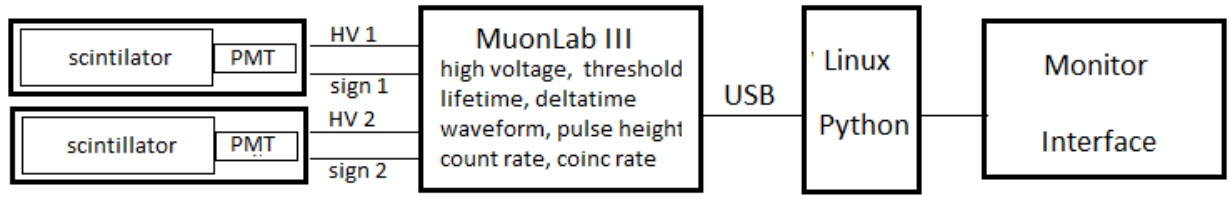

Figure 1

A detector comprises the scintillation material in the form of a rectangular block of 9 x 5 x 40 cm$^3$. The scintillator is connected to a photo-multiplier tube (PMT) and is completely enclosed in rectangular aluminum tube of 3 mm wall thickness and length of approximately 60 cm. The short light flashes generated by charged particles e.g. muons and electrons, traversing the scintillator are converted into electrical pulses by a Hamamatsu 6095 photomultiplier.

The electronics of the signal analyzer is based on an FPGA hardware circuit and a VHDL firmware program, which is saved in an EPROM. The hardware uses a 1000 MHz clock and the implemented code offers a number of possibilities for measurements of muons and electrons. Control and read-out can be realized via USB connection to a Linux computer.

We used the capability of the signal analyzer to measure the time difference (deltatime) between the pulses triggers in the two detectors at input channel 1 (CH1) and input channel 2 (CH2). Based on the 1000 MHz clock, the firmware offers the possibility of measuring time differences with a resolution of 0,5 ns.
For the measurement of pulse heights the signal analyzer contains a 200 MHz Analog to Digital converter (ADC) with a maximum range of 2 V.

High voltages for the photomultipliers and thresholds for the trigger levels can be set using an external Python program.

The firmware of the signal analyzer distinguishes between the first triggers in CH1 and CH2. In the control software, we add a positive sign to a deltatime arising from the first trigger in detector1 at CH1, whereas a deltatime arising from the first trigger in detector2 at CH2 is counted as negative. Accordingly, positive values of deltatime correspond to situations in which detector1 had the first hit and negative values of deltatime correspond to situations in which detector2 was hit first.

We performed several measurements with the detectors in two different arrangements and at various distances. In the vertical arrangement the detectors are placed parallel above each other, in the horizontal arrangement parallel next to each other. Because of the high count rates resulting from a high photomultiplier (PMT) amplification factor and a low trigger level, we used a dedicated Python program for data collection. The originally used LabView program appeared unsuitable for high count rates. The results are presented in the form of histograms, showing the vents within a time interval -50 ns to +50 ns and scaled to a measurement time of 24 h.

## 3. Results deltatime

All measurements were scaled to a measurement time of 24 h.
For convenience indicator lines at -10 ns, 0 ns and +10 ns are drawn in the histogram plots.

3.1 Vertical arrangement, detectors above each other at 2m distance, effect of PMT high-voltage

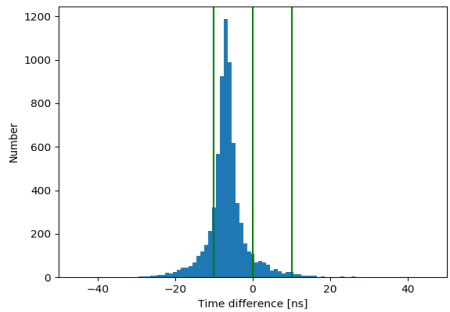

Figure 2
2m vertical HV 800 V, trigger level 40 mV

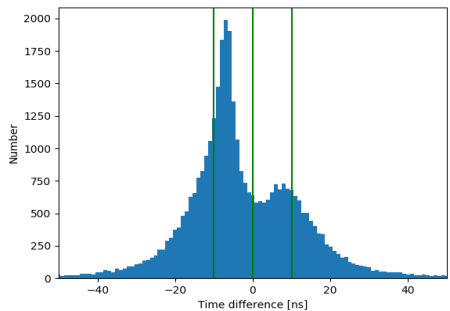

Figure 3
2m vertical HV 1100 V, trigger level 40 mV

We observe a large peak at a negative value of deltatime. It is obvious that this peak corresponds to the time of flight of a muon from upper detector2 to lower detector1, after the first hit in the upper detector. After we increased the amplification by choosing a larger value of the PMT high voltage, a second, smaller peak was observed at a positive value of deltatime. The position of the small peak corresponds to the flight time, which can be calculated from a velocity that is approximately equal to the speed of light. Therefore, it seems that the small peak originates from particles travelling at approximately the speed of light from detector1 to detector2, after a first hit in the lower detector.

To further investigate this effect, we continued our experiment in a horizontal arrangement in which the two detectors were placed parallel to each other at a mutual distance of 2m.

3.2 Horizontal arrangement, detectors next to each other at 2m distance, effect of PMT high-voltage.

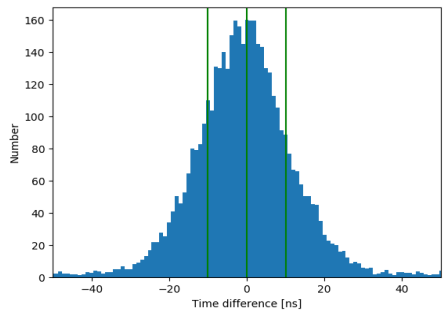

Figure 4
2m horizontal HV 800 V, trigger level 10 mV

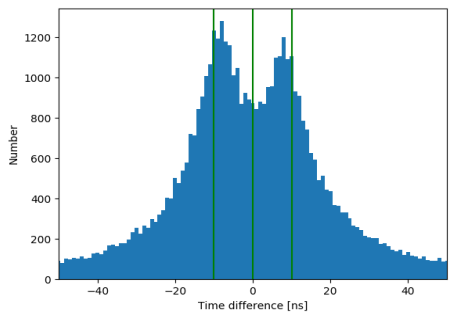

Figure 5
2m horizontal HV 1100 V, trigger level 10 mV

At higher values of the PMT voltage we observe the appearance of 2 peaks, indicating that in the vertical arrangement a small peak at a negative deltatime was hidden by the large muon time-of-flight peak.

3.3 Effect of lead halfway between the detectors

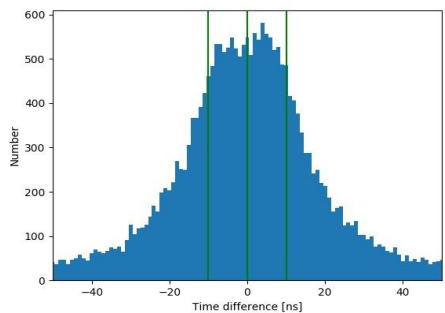

Figure 6
2m horizontal HV 1100 V, trigger level 10 mV
4 mm lead halfway between detectors

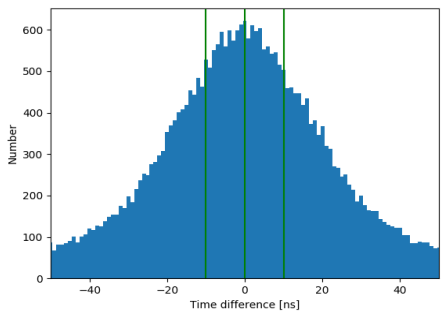

Figure 7
2m horizontal HV 1100 V, trigger level 10 mV
10 mm lead halfway between detectors

Apparently the separate peaks, presumably resulting from transfer of information between the two detectors, can be blocked by lead at a position halfway between the detectors.
With 4 mm of lead, we still observe in Figure 6 the remains of a structure. Further measurements at 2mm, 6mm and 8mm (results not shown) indicate that the disappearance is gradual. Under the conditions of HV 1100 V and 10 mV threshold, 10 mm of lead are required for the two peak structure to completely vanish.

 It appears that there is no influence from lead plates placed at the ouside of the detectors.

3.4  Horizontal arrangement, detectors parallel to each other, effect of distance between detectors.

We made measurements at 2m, 2,5m, 3m, 3,5m, and 4m, examples are shown in Figure 8, Figure 9 and Figure 10

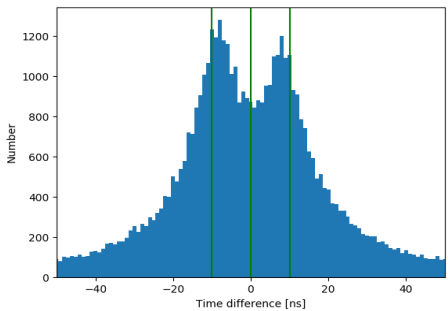

Figure 8
2m horizontal HV 1100 V, trigger level 10 mV

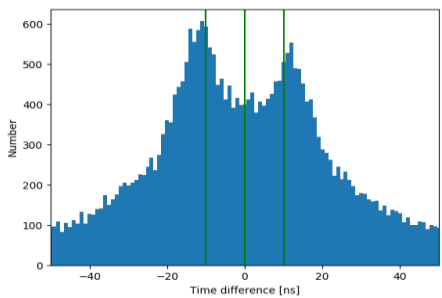

Figure 9
3m horizontal HV 1100 V, trigger level 10 mV

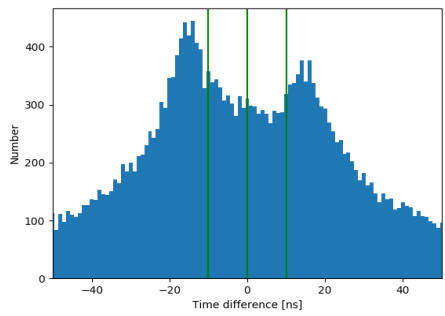

Figure 10

4m horizontal HV 1100 V, trigger level 10 mV

We observe a shift in the position of the peak that results from the first hits in CH1 as well as in the position of the peak that results from the first hits in CH2. The asymmetry in the plots is presumably owing to different sensitivities of the detector channels.

## 4. Analysis of the deltatime results

We analyzed the deltatime data, with all files scaled to a measurement time of 24 h, in the interval from -150 ns to +150 ns. At all distances between the detectors we made a least-squares fit based on the assumption that the observed histogram is the result of a constant background plus three Gaussian peaks. From the analysis we obtained the positions of the peaks, maximum amplitudes, widths and vertical offset resulting from the background. The graphical results for three examples of the analyses at distances of 2m , 3m and 4m between the detectors are shown in Figures 11, 12 and 13. From visual judgment of the graphs we can conclude that the fit to the observed data is very good. Least-squares analyses have not been performed, but are believed to confirm the observations.

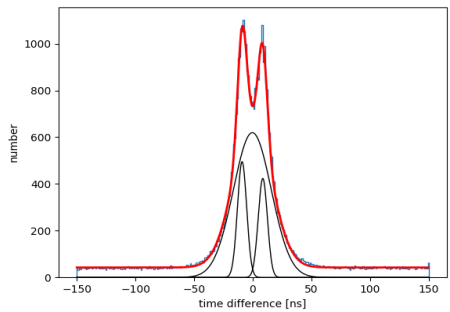

Figure 11

2m horizontal, HV 1100 V, trigger level 10 mV

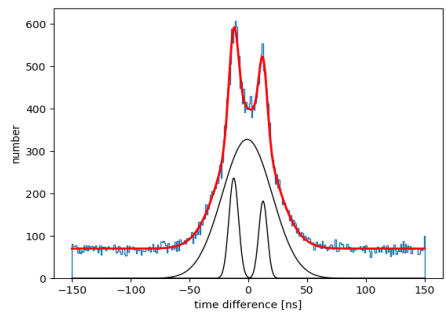

Figure 12

3m horizontal, HV 1100 V, trigger level 10 mV

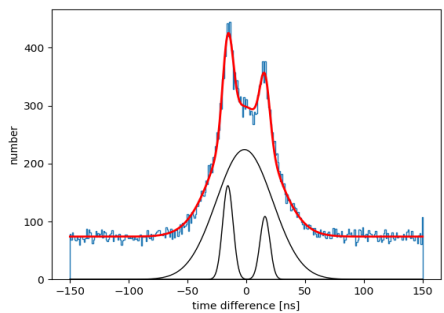

Figure 13

4m horizontal, HV 1100 V, trigger level 10 mV

Quantitative results for the number of coincidences within the three peaks and in a uniform background:

Table 1

|  |  | 2m | 2,5m | 3m | 3,5m | 4m |
|---|---|---|---|---|---|---|
| offset | background | 43 | 58 | 70 | 67 | 73 |
| middle peak | max | 619 | 458 | 328 | 287 | 224 |
|  | pos (ns) | -0,4 | -0,4 | -1 | -0,8 | -1,6 |
|  | width (ns) | 32,4 | 39,2 | 42,4 | 44,4 | 47,8 |
|  | counts | 25077 | 22511 | 17429 | 15935 | 13475 |
| left peak | max | 495 | 358 | 237 | 201 | 160 |
|  | pos (ns) | -9,1 | -10,6 | -12,4 | -14 | -15,7 |
|  | width (ns) | 7,8 | 8,6 | 8,2 | 8,0 | 9,0 |
|  | counts | 4871 | 3845 | 2421 | 2019 | 1787 |
| right peak | max | 424 | 310 | 182 | 144 | 108 |
|  | pos (ns) | 8,5 | 10,3 | 12,6 | 13,5 | 16 |
|  | width (ns) | 7,8 | 7,6 | 7,4 | 6,4 | 8,4 |
|  | counts | 4137 | 2961 | 1668 | 1169 | 1125 |

The constant offset (plateau) is due to random coincidences which are uniformly distributed in time. We may estimate the number of random coincidences per bin of 1 ns from the count rates of the two detectors, which varied during the measurements around 850 c/s with a variation of about 5 %. This leads to a background per 1 ns bin of
$850 \times 850 \times 10^{-9} \times 24 \times 3600$ = approximately 60 random events per bin in 24 hours, in reasonable agreement with the observation.

We interpret the middle peak, centered around zero deltatime, as originating from extended air showers (EAS), coming from all directions and at all possible angles. The angle distribution of the wave fronts causes a spread in deltatime around zero.

For the two small peaks, centered around a positive and a negative value of deltatime, we have no interpretation. We conclude that, after a first hit in CH1, (deltatime counted positive) or in CH2, (deltatime counted negative), information is transferred between the detectors in a globally symmetric manner, giving rise to a coincidence.

Apart from an offset, which may be due to an electronic delay in the hardware, the velocity of information transfer is close to the speed of light, as can be seen from a plot of half of the separation of the two peaks against the distance of the detectors:

| Distance between the detectors | 2m | 2,5m | 3m | 3,5m | 4m |
|---|---|---|---|---|---|
| separation DT between peaks / 2 (ns) | 8,80 | 10,45 | 12,50 | 13,75 | 15,85 |

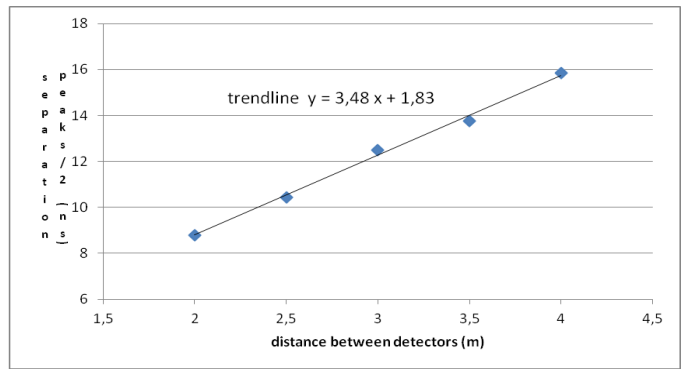

Figure 14
Plot of half peak separation in ns versus distance of detectors in m
For comparison: the inverse of the speed of light is 3,34 ns/m

From the fitted data, we considered the number of events in the two small peaks.

| Distance between the detectors | 2m | 2,5m | 3m | 3,5m | 4m |
|---|---|---|---|---|---|
| Counts left peak | 4871 | 3845 | 2421 | 2019 | 1787 |
| Counts right peak | 4137 | 2961 | 1668 | 1169 | 1125 |

A plot of the number of events against distance suggests a quadratic inverse power law dependence on distance. These results suggest that the detector material is a source of (secondary?) radiation, which is visible only in the region of low-amplitude signals.

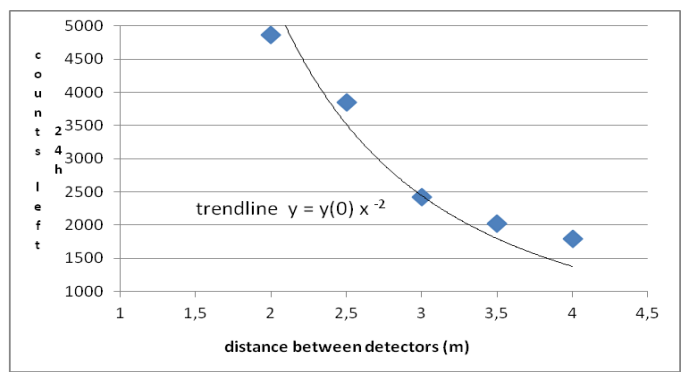

Figure 15
Counts left peak (CH2 first) against distance detectors compared with an inverse quadratic dependence

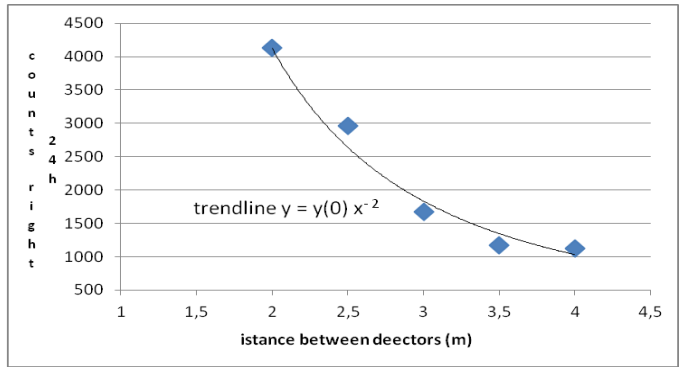

Figure 16

Counts right peak (CH2 first) against distance detectors compared to an inverse quadratic dependence

In contrast to the behavior of the left and right peaks, the counts in the middle peak showed a linear dependence on distance between the detectors.

| Distance between the detectors | 2m | 2,5m | 3m | 3,5m | 4m |
|---|---|---|---|---|---|
| Counts middle peak | 25077 | 22511 | 17429 | 15935 | 13475 |

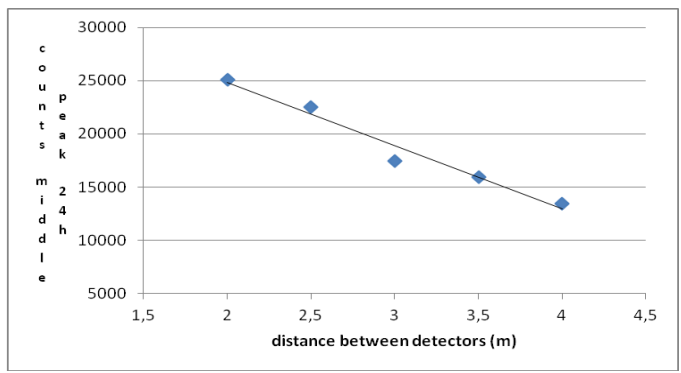

Figure 17

Number of counts middle peak against distance of detectors

## 5. Effect of trigger level

At a horizontal distance of 2m we made further measurements using various values of the trigger level, which were scaled to a measurement time of 24 h. We made Gaussian fits between -100 ns and +100 ns, the result of which are shown in Figures 18, 19 and 20. Comparing Figure 18 to Figure 11, we noticed a considerable quantitative (statistical?) difference. However, the effect remained qualitatively the same.

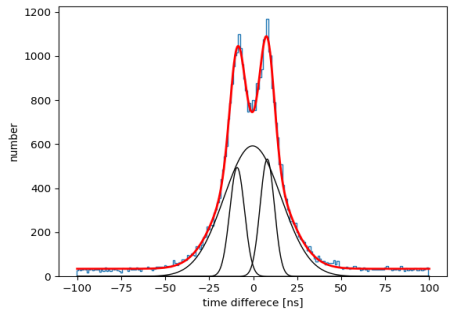

Figure 18
2m, HV 1100 V, trigger level 10 mV

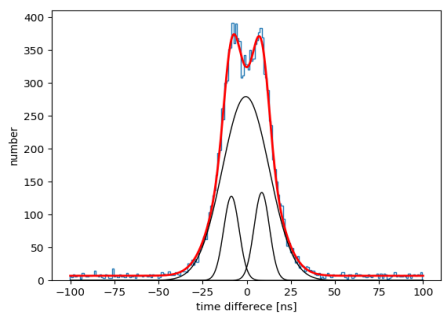

Figure 19
2m, HV 1100 V, trigger level 70 mV

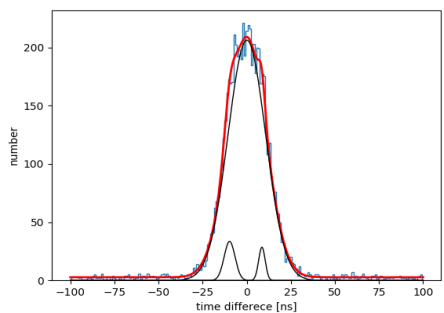

Figure 20
2m, HV 1100 V, trigger level 150 mV

Resulting from the Gaussian fits:

| trigger level | 10 mV | 70 mV | 150 mV |
|---|---|---|---|
| middle counts | 24062 | 9325 | 5518 |
| left counts | 5193 | 1407 | 537 |
| right counts | 5387 | 1462 | 259 |
| (le + ri) / mid | 44 % | 31 % | 14 % |

We observe that at higher values of the trigger level the two-peak structure in the deltatime spectrum becomes less prominent. At higher values of the trigger level finally the peak structure completely vanishes.

6. Pulse height

To shed more light on the nature of the supposed information transfer between the two detectors, we performed separate measurements on the pulse height of the signals at different trigger levels.

Using the ADC capability of the analyzer box, we registered the amplitude of the signal in CH1 in a separate file PH_CH1, under the condition that there is a trigger from a coincidence between CH1 and CH2 within a fixed time window of 100 ns. Together with the CH1 signal amplitude, the file registers the deltatime between CH1 and CH2 and the information on which channel was hit first. The ADC only measures the signal at CH1 input. After each coincidence, the maximum amplitude is stored in a 255 bits memory and, together with the information on the corresponding deltatime, written to the datalog file PH_CH1. The analog range of the ADC is 0-2000 mV; therefore, one ADC bit corresponds to an analog voltage of approximately 8 mV and the analog pulse height PH equals ADC bits*8 mV

From the files PH_CH1, we separated the counts in the bit range 0 - 255 and the counts collected at the maximum bit 255 of the ADC. As shown above, these ranges correspond to the voltages of pulse height 0<PH<2000 mV and PH>2000 mV.

ADC counts in the region DT 0 to +50 ns (CH1 first)

|  | 0<PH<2000 mV | PH>2000 mV |
|---|---|---|
| trigger level 10 mV | 14153 | 2357 |
| trigger level 70 mV | 4356 | 1445 |
| trigger level 150 mV | 1571 | 1030 |

ADC counts in the region DT -50 to 0 ns (CH2 first)

|  | 0<PH<2000 mV | PH>2000 mV |
|---|---|---|
| trigger level 10 mV | 16585 | 468 |
| trigger level 70 mV | 5302 | 429 |
| trigger level 150 mV | 2323 | 410 |

Globally, we noticed an equal number of total counts in the CH1 first region and the CH2 first region, in accordance with the observed global symmetry in the deltatime spectrum. Because the ADC only measures the signal in CH1, the number of PH>2000 mV at CH1 first is larger than at CH2 first.

We created histograms of the number of pulse heights at signal analyzer input 1 with respect to amplitude for the two regions of deltatime.

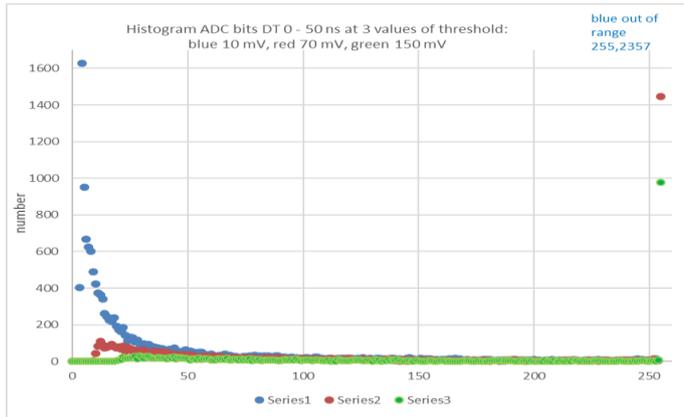

Figure 21

Histograms of ADC counts (PH) In the CH1 first region

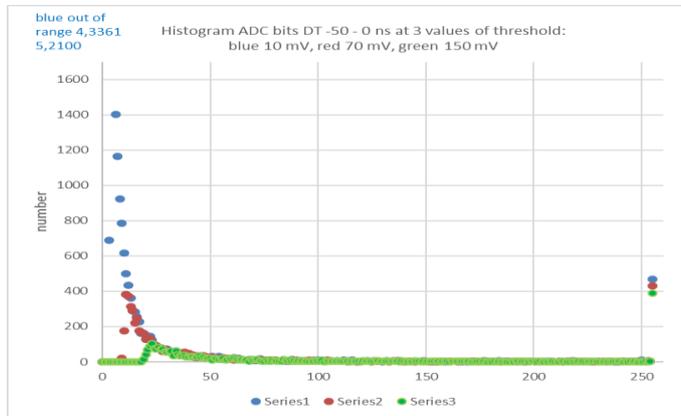

Figure 22

Histograms of ADC counts (PH) In the CH2 first region

We observed a large contribution at small ADC bit values, with maxima in the ADC bit region corresponding to the chosen trigger voltage. The number of counts at low ADC bit values almost disappeared at the highest chosen trigger level.

These observations, in combination with the vanishing of the two-peak structure at higher values of the trigger level in the Gaussian fits, lead us to conclude that the peaks in the deltatime spectrum are associated with small values of the pulse height PH, measured by the ADC at signal analyzer input 1.

The energy absorption associated with small values of the pulse height can be roughly estimated as follows:
A zero value of the pulse height implies a signal amplitude smaller than one ADC bit of 8 mV. Assuming that in a scintillator with a thickness of 5 cm the maximum signal amplitude corresponds to an absorbed energy of the order of 10 MeV (i.e. 5 MIP), we estimate that at a PMT voltage of 1100 V a signal amplitude of the order of 10 mV of the ADC corresponds to an absorbed energy of approximately 50 keV. From the histograms, we see that the disappearance of the two peak structures, as shown in Figures 18, 19 and 20 has to be associated with ADC bits smaller than 25 (corresponding to a threshold of 200 mV). According to our estimates, these values correspond to absorbed energies smaller than 1 MeV. The histogram at 10 mV trigger indicates that the peak structure in the deltatime spectrum is related to much smaller energy absorptions, on the order of 100 keV

## 7. Discussion

We observed that, both in the vertical and in the horizontal arrangement of two scintillation detectors deltatime measurements show unexpected coincidences between signals at low amplitudes in both detectors.

From the least-squares Gaussian analysis of the deltatime histogram we conclude that the observed histogram is the result of a constant background plus three Gaussian peaks. One peak is centered around DT = 0 ns, the other two peaks are centered around DT = -8 ns and DT = +8 ns.

As a first hypothesis, we assume that the middle peak in the least-squares Gaussian fit results from coincidences associated with wave fronts of an EAS shower coming either from the left or right. For spatial arguments, we expect the fit result for the middle peak to be symmetrical around zero deltatime.

The origin of the other two peaks is not clear. The position of the peaks suggests that there might be a second source of coincidences, connected by a mutual transfer of information between the detectors at a velocity close to the speed of light. This assumption is confirmed by measurements of the deltatime spectrum as a function of the distance between the detectors. In addition, measurements of the number of events at various distances suggest an approximately inverse quadratic dependence, which could indicate that the detectors act as a spherical symmetric source of radiation.

Separate measurements of the deltatime spectrum at various trigger levels and pulse height measurements using an ADC at one of the scintillation detector inputs showed that the two extra peaks in the spectrum are caused by signals of very small amplitude. If related to energy absorption in the scintillator, the absorbed energy was on the order of 100 keV.

For an explanation we considered two possibilities:
First, Compton scattering of air shower photons in the scintillator might be involved. In this hypothesis we must assume that scattered photons escape from one detector scintillator and travel to the other detector. This could be the reason for the extremely small amplitudes of the signals in the second detector.
Second, environmental radioactivity[4], like radon from concrete walls, might have contributed to the deltatime data. This also could result in small amplitude signals.

Considering one of the scintillators as a source radiating uniformly in all directions and the other scintillator as a flat 200 $cm^2$ detector at a distance of 2m, we globally calculated from the actual measured number of counts in the left and right peak, the apparent strength of the source. The calculation leads to a strength of approximately 150 source events/s, which is more than 10 times higher than that expected from intensity measurements on cosmic air showers (see ref 3) or from environmental radioactivity (see ref 4).

Therefore, both hypotheses were rejected.

The diminishing and final disappearance of the left and right peaks after the introduction of lead between the detectors, suggests that particles transferring information, for example electrons or photons, might be involved in the information transport. These particles must originate from the detectors because lead plates only have influence if they are placed between the detectors.
However, the observation of remains of a structure in Figure 6 at a lead thickness of 4 mm is contradictory to our conclusion from trigger level and pulse height measurements that the energies involved were in the order of 100 keV. Half-value thicknesses of lead for photons at 100 keV and 200 keV are 0,12 mm and 0,68 mm respectively. For electrons of 200 keV the penetration depth in lead is in the order of 0,1 mm.

Apart from this, from the effect of disappearance of the peaks after the introduction of lead between the detectors, we may conclude that we do not deal with an instrumental effect of the signal analyzer.

## 8. Conclusion

The measurement of time differences between signals in two scintillation detectors in a horizontal arrangement at high values of the PMT voltage and at low values of the trigger level clearly shows a two-peak structure in the deltatime spectrum at low signal amplitudes.

A least-squares fit with three Gaussians first suggests that air showers from cosmic rays are involved in the production of the signal pulses and second indicates that information is transferred between the detectors after a trigger in one of the detectors.

Usually, measurements of cosmic radiation are made with PMT high voltages and signal trigger levels that are chosen to avoid the influence of noise. Consequently, phenomena at low signal amplitudes are hidden, do not become visible in the spectra and are interpreted as noise. However, measurement of the time spectrum of coincidences appears to be a sensitive way to detect the presence of unexpected phenomena and to reveal a structure in the noise.
To date, the origin of this structure is unclear.

It is interesting to note that the coincidence method applied to the situation of high PMT voltage and low trigger level might be able to reveal additional information on the interaction process in the detector material. Surprisingly, the information only becomes visible clearly in a situation where a high noise level is allowed.

The MuonLab apparatus, designed for projects at the high school and academic bachelor levels, offers two possibilities for detecting fundamental phenomena in cosmic ray physics. First, the muon lifetime and muon velocity can be measured in the vertical setting of the detectors. Second, coincidences produced by secondary particles in material above the detectors (Rossi effect) and, moreover, up to now unexplained coincidences at low signal level can be measured in the horizontal setting of the detectors.

Further experiments on the peak structure in the deltatime spectrum are required to reveal the mechanism of the coincidences and to possibly provide an explanation.

## Acknowledgements

The measurements I carried out were performed within the Hisparc Group of the Nikhef National Institute for Subatomic Physics. First, I would like to thank the Board of the Nikhef Institute and Professor Frank Linde for their support . For many years I have experienced a very healthy research atmosphere at the Institute. I thank Professor Bob van Eijk, head of the Hisparc group, for giving me the opportunity to work as a guest collaborator in his Hisparc laboratory. I am very grateful to Kasper van Dam and Jos Steiger for stimulating discussions and for their valuable contribution as experienced scientists in the analysis of the data. In addition, I would like to thank Hans Verkooijen, Electronic Engineer at the Institute and designer of the MuonLab signal analyzer, for always being available for information and modification.
Warm thanks all of you.